\documentstyle[aps,epsfig,12pt]{revtex}

\newcommand{\mc}{\;\mbox{,}}

\begin{document}
\title{{\bf{ Superconductivity in a two-component model with local electron
pairs}}}
\author{ R. Micnas$^{1}$, S. Robaszkiewicz$^{1}$ and A. Bussmann-Holder $^{2}$}

\address{$^1$Institute of Physics,  A. Mickiewicz  University,
             Umultowska 85, 61-614 Pozna\'{n}, Poland }
\address{$^2$ Max-Planck Institut f\"ur Festk\"orperforschung, Heisenbergstrasse1, D-70569 Stuttgart,
Germany}
\date{\today}
\maketitle

\begin{abstract}
Superconductivity in the two component model  
 of  coexisting 
local electron pairs (hard-core charged bosons) and itinerant fermions  coupled 
via charge exchange mechanism
is discussed.
The cases of isotropic $s$-wave and anisotropic pairing of extended $s$-wave and
$d_{x^2-y^2}$   symmetries are   analyzed
for a 2D square lattice  within the BCS-mean field approximation 
and the Kosterlitz-Thouless theory.
The phase diagrams 
and superconducting characteristics of this induced pairing model as a function of 
the position of the local pair (LP) level and the total
carrier concentration are determined. The model exhibits several types of interesting 
crossovers between the  BCS like behavior
and that of  LP's.  In addition, the Uemura plots are  obtained for extended $s$ and 
$d_{x^2-y^2}$ pairing symmetries.
Finally, we analyze
the  pairing fluctuation effects (in 3D) within a generalized 
$T$-matrix approach. 
Some of our results are discussed in 
connection with  a two-component scenario 
of preformed pairs and unpaired electrons 
for  high temperature
superconductors.
\end{abstract}


%\end{keyword}

%%%%%%%%%%%%%%%%%%%%%%%%%%%%%%%%%%%%%%%%%%%%%%%%%%%%%%%%%%%%%%%%%%%%%%%%
%%                     
%%%%%%%%%%%%%%%%%%%%%%%%%%%%%%%%%%%%%%%%%%%%%%%%%%%%%%%%%%%%%%%%%%%%%%%%
%\newpage
\section{Introduction}
A mixture 
of interacting charged bosons 
  (local electron pairs with $q=2e$)  and  electrons  can  show  features  which  are 
  intermediate between those of  local pair (bipolaronic)   superconductors and those 
  of classical BCS systems. Such a  two component (boson-fermion) model is of relevance
  for high temperature superconductors (HTS) and other exotic superconductors
\cite{{Micnas90},{gorkov87},{rob87},{lee89},{jrmr96},{NATO},{larkin97},{czart01},{castro},{altman},{rmsrbt},{unpublished},{physicac}}.
A similar model has also been adopted  for the description of
a resonance
 s-wave superfluidity  in Fermi atomic gases with a Feshbach resonance
  \cite{{holland},{ohashi}}. \\ Recently, we have studied a generalization of this model to the case of
  anisotropic pairing \cite{rmsrbt,unpublished,physicac}. Here, we briefly outline 
 the study 
 and  present some further results concerning the phase diagrams
  and superconducting properties of such a system  in the 
  case of isotropic $s$-wave 
  and  anisotropic $d$-wave pairing, 
  for a 2D square lattice (Sec.II-III), 
  as well as for s-wave pairing for a 3D simple cubic (sc) lattice (Sec.IV).
 
 \section{The Model}
 We consider the model of coexisting
 electron pairs (hard-core bosons "b") and itinerant $"c"$ electrons defined by 
 the following effective   Hamiltonian 
\begin{eqnarray}\label{main}
{\it H} =  \sum_{\bf {k}\sigma} (\epsilon_{\bf k}-
\mu)c^{\dagger}_{\bf{k}\sigma}
c_{\bf {k}\sigma} + 2\sum_{i}(\Delta_{0}-\mu)b^{\dagger}_{i}b_{i}
-\sum_{ij}J_{ij}b^{\dagger}_{i}b_{j}
+\nonumber \\
\sum_{\bf k, q}\left[V_{{\bf q}}({\bf k}) c^{\dagger}_{\bf{k} +\bf {q}/2,
\uparrow}c^{\dagger}_{{\bf {-k} +\bf {q}}/2, \downarrow} b_{\bf {q}} 
+ h.c.\right] +{\tilde H}_{C},
\end{eqnarray}
where $\epsilon_{\bf k}$  refers to  the   band energy of  the  c-electrons, 
  $\Delta_{0}$   measures  the 
  relative position of the LP level with respect to the bottom of 
  the c-electron band, 
  $\mu$ is the chemical potential which ensures that the 
  total number of particles in the system is constant, i.e. 
% \begin{eqnarray}\label{numbereqn}
$ n=\frac{1}{N}\left(\sum_{\bf k\sigma}
\langle c^{\dagger}_{\bf k \sigma}c_{\bf k \sigma} \rangle
+2\sum_{i} \langle b^{\dagger}_{i}b_{i} \rangle\right)=n_{c}+2n_{B}.$
%\end{eqnarray}
$n_{c}$ is the concentration of $c$-electrons, $n_{B}$
is the number of
 local pairs  per site. $J_{ij}$ is the pair hopping integral.
${\tilde H}_{C}$ denotes Coulomb interaction terms. 
The  
operators for local pairs 
$ {b^{\dagger}_{i},b_{i}}$ 
obey the Pauli spin 1/2 commutation rules. $V_{\bf q}(\bf k)$ describes the coupling
between the two subsystems.  We  will consider the case 
$V_{{\bf q}}({\bf k})=V_{0}({\bf k})= I \phi_{k}/\sqrt{N}$, 
and neglect its $q$ dependence  at small $q$. 
The interaction term
takes  the form of coupling,   
via the center of mass momenta ${\bf q}$,  of the singlet pair of $c$-electrons
$B^{\dagger}_{\bf q}$
 and the hard-core boson $b_{\bf q}$:
 \begin{eqnarray}
H_{1}= \frac{1}{\sqrt{N}}\sum_{\bf q}I(B^{\dagger}_{\bf q}b_{\bf q} 
+b^{\dagger}_{\bf q}B_{\bf q}).
\end{eqnarray}
$B^{\dagger}_{\bf q}=\sum_{\bf k}\phi_{k}c^{\dagger}_{\bf{k} +\bf {q}/2,
 \uparrow}c^{\dagger}_{\bf {-k} +\bf {q}/2, \downarrow}$ denotes 
 the singlet pair creation
operator of $c$-electrons and $I$ is the coupling constant.
 The pairing symmetry, on a 2D square lattice,  is determined by the form of 
 $\phi_{k}$, 
 which is constant ($1$)  for  on-site pairing ($s$),
 $\phi_{k}= \gamma_{k}=\cos(k_{x})+\cos(k_{y})$ for  extended $s$-wave
 ($s^{*}$) and $\phi_{k}= \eta_{k}=\cos(k_{x})-\cos(k_{y})$ for  
 $d_{x^2-y^2}$-wave pairing ($d$).
In general, one can consider a decomposition 
 $I\phi_{k}=g_{0}+g_{s}\gamma_{k}+g_{d}\eta_{k}$, 
with appropriate coupling parameters for different 
 symmetry channels.

 %For the system under consideration we define 
 The superconducting state of the model is characterized by  two 
 % superconducting  
 order parameters: $x_{0}=
\frac{1}{N}\sum_{k}\phi_{k} \langle
c^{\dagger}_{k\uparrow}c^{\dagger}_{-k\downarrow} \rangle$
 and $\rho^{x}_{0}=\frac{1}{2N}\sum_{i}\langle b^{\dagger}_{i}+b_{i} \rangle$.
In the BCS-mean-field approximation (MFA)  
the free energy of the system (for ${\tilde H}_{C}=0$) is evaluated to be:
\begin{eqnarray}\label{freeenergy}
F/N = -\frac{2}{\beta N}\sum_{{\bf k}}\ln\left[2\cosh(\beta E_{{\bf
k}}/2)\right]
-\frac{1}{\beta}\ln\left[2\cosh(\beta\Delta)\right] + C ,\\
C = -\epsilon_{b} +\Delta_{0}+\mu (n_{c}+2n_{B})-2\mu -2I |x_{0}|\rho_{0}^{x}+J_{0}(\rho^{x}_{0})^2~,
\end{eqnarray}
where the quasiparticle energy of the $c$-electron subsystem 
is given by \\$E_{\bf k}=
\sqrt{\bar\epsilon_{\bf k}^{2}+\bar{\Delta}_{\bf k}^{2}}$ and $\bar\epsilon_{\bf k}=\epsilon_{\bf k}-\mu$, 
$\bar{\Delta}_{\bf k}^2=I^2\phi_{k}^2(\rho_{0}^{x})^2$, 
$\Delta=\sqrt{(\Delta_{0}-\mu)^2+(-I|x_{0}|+J_{0}\rho^{x}_{0})^2}$. \\
$J_{0}=\sum_{i\neq j} J_{ij}$.  $\beta=1/k_{B}T.$ 
For the 2D square lattice the  $c$-electron  dispersion is  
$\epsilon_{\bf k}~=~\tilde{\epsilon}_{\bf k}-\epsilon_{b}=
-2t\left[\cos (k_{x})+\cos (k_{y})\right]-
4t_{2}\cos (k_{x})\cos (k_{y})-\epsilon_{b}$,
with the  nn and nnn hopping parameters 
$t$ and $t_{2}$, respectively, $\epsilon_{b}=min \tilde{\epsilon}_{\bf k}$ .
% $\mu$ is the chemical potential.
It should be noted that the energy gap in the $c$-band  is due 
to nonzero Bose condensate amplitude $(|\langle b \rangle |\neq 0)$,  
and well defined Bogoliubov quasiparticles can exist in the superconducting phase. 
The order parameters and the chemical potential are given by 
 \begin{eqnarray}\label{orderpar}
 \frac{\partial F}{\partial x_{0}}=0,~~ 
 \frac{\partial F}{\partial \rho_{0}^{x}}=0,~~
  \frac{\partial F}{\partial \mu}=0.
 \end{eqnarray}
The superfluid stiffness  derived within the linear response method and BCS
theory, for the case $J_{ij}=0$,
is of the form:
\begin{eqnarray}\label{stiffness}
\rho_{s}=\frac{1}{2N}\sum_{k}
\left\{
\left(\frac{\partial\epsilon_{\bf k}}{\partial k_{x}}\right)^{2}
\frac{\partial f(E_{\bf k})}{\partial E_{\bf k}}
+\frac{1}{2}\frac{\partial^{2}\epsilon_{\bf k}}{\partial k_{x}^{2}}
\left[
1-\frac{\bar\epsilon_{\bf k}}{E_{\bf k}}
\tanh\left(\frac{\beta E_{\bf k}}{2}\right)
\right]
\right\} \mc
\end{eqnarray}
%\begin{multicols}{2}
where  $f(E_{k})=1/\left[\exp(\beta E_{k})+1\right]$ is the Fermi-Dirac
distribution function. 
In the local limit: $\lambda^{-2}\propto (16\pi e^2/\hbar^2c^2)\rho_{s}$, where 
$\lambda$ is the London penetration depth.\\
The mean-field transition
temperature ($T_{c}^{MFA}$), at which the gap amplitude vanishes,
%($x_{0}\rightarrow 0,\rho_{0}^{x}\rightarrow 0$)
yields an estimation
of the $c$-electron pair formation temperature \cite{unpublished,physicac}
and is given by
\begin{eqnarray}\label{tcmfa}
1=\left[J_{0}+
\frac{I^2}{N}\sum_{\bf k}
\phi^2_{k}\frac{\tanh\left(\beta_{c}^{MFA}\bar\epsilon_{\bf k}/2\right)}
{2\bar\epsilon_{\bf k}}\right]
\frac{\tanh\left[\beta_{c}^{MFA}(\Delta_{0}-\mu)\right]}{2(\Delta_{0}-\mu)}.
\end{eqnarray}

Due to the fluctuation effects the superconducting
phase transition will occur at a critical temperature  lower
than that given by the BCS-MFA theory.  In 2D, $T_{c}$ can be derived 
within the Kosterlitz-Thouless (KT) theory for 2D superfluids 
 \cite{KT}, which describes the transition 
in terms of vortex-antivortex
pair unbinding.    We evaluate $T_{c}$
 using the KT relation for the universal jump of
the (in-plane) superfluid density $\rho_{s}$ at $T_{c}$ \cite{KT} :
\begin{equation}
\frac{2}{\pi}k_{B}T_{c}=\rho_{s}(T_{c}),
\end{equation}
where $\rho_{s}(T)$ is given by Eq.(\ref{stiffness}) and  $x_{0}(T), 
\rho_{0}^{x}(T), \mu (T)$ are given by 
Eqs.(\ref{orderpar}). 
%Thus, 
Thus, the critical temperature denoted further by $T_{c}^{KT}$ is determined from the
set of four self-consistent equations. In the weak coupling limit 
$(|I_{0}|/2D\ll 1, J_{0}=0, D$-the half-bandwidth, $I=-|I_{0}|$), $T_{c}^{KT}/T_{c}^{MFA}\rightarrow 1$
if  $|I_{0}|/2D\rightarrow 0$.

\section{Results for 2D electron spectrum}
A comprehensive 
%an extended 
analysis of the phase diagrams and 
superfluid properties of the model Eq.(\ref{main}) for different pairing
symmetries including $s$, the extended $s$ ($s^{*}$) and $d_{x^2-y^2}$-wave
symmetries was performed in Refs.\cite{rmsrbt,unpublished,physicac}. 
Below we  will discuss  
these results including some
additional ones. (The term ${\tilde H}_{C}$ will not be considered).

In the absence
  of interactions, 
depending on the relative concentration of "c"  electrons and LP's
  we distinguish three essentially different physical situations. For $n\leq2$ it  will be:\\ 
  (i) $\Delta_{0} <0$ such that at $T=0K$ all the available electrons form 
local pairs  ($2n_{B}\gg n_{c}$) (LP); \\
  (ii) $\Delta_{0} >0$ such that the "c" electron band 
  is filled  up  to  the 
  Fermi 
  level $\mu=\Delta_{0}$  and the remaining electrons are in the 
  form of  local pairs  (the  "c+b" or Mixed regime, $0<2n_{B},n_{c}<2)$
  (LP+E);\\ 
  (iii) $\Delta_{0} >0$ such that the Fermi level  $\mu<\Delta_{0}$ 
    and  consequently  at $T=0$K all  the 
  available electrons occupy the "c" electron states (the c-regime or"BCS",
$n_{c}\gg 2n_{B}$) (E).

For  $|I_{0}|\neq 0$,  in  the  case  (ii)  superconductivity  
is due to the  
  interchange between local pairs  and  pairs  of  "c" 
  electrons. In this process "c" electrons become "polarized" into Cooper 
  pairs and local pairs increase their mobility by  decaying  into 
  "c" electron pairs. In this intermediate case neither the  standard  BCS 
  picture nor the picture of local pairs  applies  and  superconductivity 
  has a "mixed" character.
  The system shows features which are intermediate between the BCS and
preformed local pair regime. This concerns the energy 
gap in the single-electron
excitation spectrum ($E_{g}(0))$, the $k_{B}T_{c}/E_{g}(0)$ ratio, the 
critical fields,
the Ginzburg ratio $\kappa$, the width of the critical regime 
 as well as the normal state properties.
  In case (i) the local pairs can move  
  via  a mechanism of virtual 
  excitations into empty c-electrons states. Such a mechanism  gives  rise 
  to the long range hopping of  LP's 
 (in analogy  to the  RKKY 
  interaction  for s-d mechanism in the magnetic equivalent).  
The superconducting properties are analogous to those of  a pure local 
pair (bipolaronic) superconductor  \cite{{Micnas90},{NATO},{ASA}}.
In  case  (iii),  on  the 
  contrary, we find a situation which is similar to the BCS case: pairs of 
  "c" electrons with opposite momenta and spins are exchanged via virtual 
  transitions into local pair states.
  
The generic phase diagrams for $s$-wave  pairing  symmetry
plotted as a function 
of the position of the LP level $\Delta_{0}$ at fixed $n$ 
are shown in Fig.1. In Fig.2. we show the transition temperatures for the $d_{x^2-y^2}$-symmetry.

In all the cases one observes a 
 drop in the superfluid stiffness (and in the  KT transition 
temperature) when the bosonic level reaches the bottom 
of the $c$-electron band and the system approaches the LP limit.  
In the opposite, BCS like limit, $T_{c}^{KT}$ approaches
asymptotically 
%the mean-field (MF)  transition temperature 
$T^{MFA}_{c}$,
 with  a narrow fluctuation regime.
Between the KT and MFA temperatures,   {\it phase fluctuation effects} 
are important. 
In this regime a pseudogap in the c-electron spectrum will develop and 
the normal state of LP and itinerant fermions  
can exhibit non-Fermi liquid properties \cite{jrmr96}.

A closer inspection of the Mixed-LP crossover indicates that  when the LP 
level is lowered and  reaches the bottom of the fermionic band 
an effective attraction between fermions becomes 
strong, since it varies as $I^2/(2\Delta_{0}-2\mu)$ and $\mu\approx \Delta_{0}$\cite {{unpublished},{physicac}}. 
In this regime the density of $c$ electrons is low and 
 formation of bound
$c$-electron pairs occurs.  
It gives rise 
to an energy gap  in the single-electron spectrum {\it independently of the pairing
symmetry}.
%.
 We  calculated the binding energies of $c$- 
electron pairs and
found that $T_{c}^{MFA}$ essentially scales with the  half of their binding energy for 
$\Delta_{0}< 0$.
The superconducting transition temperature  is here always 
much lower than the $c$-pair formation temperature ($T_{c}^{MFA}$) and
%quickly 
decreases rapidly with $|\Delta_{0}/D|$. In such a case, the superconducting
state can be  formed by two types of coexisting  bosons: 
preformed $c$- electron pairs and LP's \cite{unpublished,ohashi}.

Comparing $T_{c}$ {\it vs} $\Delta_{0}$ plots for various pairing symmetries  
one finds that in
the case of nn hopping only, the $d$ and $s$ -wave pairings are favorable for
higher concentration of $c$-electrons, while the $s^{*}$-wave can be stable at
low $n_{c}$.  
The nnn hopping $t_{2}$ (with opposite sign to $t$) can strongly enhance $T_{c}$
for $d$-wave symmetry, moreover it favors the $d$ and $s$-wave pairings for
lower values of $n_{c}$ (compare Fig.5) \cite{unpublished}.

The region between $T_{c}^{MFA}$ and $T_{c}^{KT}$, where the system can exhibit a
pseudogap, expands with increasing intersubsystem coupling $|I_{0}|$. 
As we have found \cite{physicac}, except for $|I_{0}|/D\ll 1$ the coupling dependences
of $T_{c}^{MFA}$ and $T_{c}^{KT}$ are qualitatively different.
 $T_{c}^{MFA}$ is an increasing function of $|I_{0}|$ for all the pairing
 symmetries. On the other hand, $T_{c}^{KT}$ {\it vs} $|I_{0}|$ increases first, goes
 through a round maximum and then decreases (similarly as it is observed in the
 attractive Hubbard model). The  position of the maximum 
 corresponds to  the intermediate
 values of $|I_{0}|/D$ and it depends on the pairing symmetry as well as the values
 of $\Delta_{0}/D$ and $n$. For large  $|I_{0}|$, the 
 %values of 
 $T_{c}^{KT}$ are close
 to the upper bound for the phase
 ordering temperature which is given by $\pi\rho_{s}(0)/2$.

Concerning  the evolution of  superconducting properties with increasing $n$
at fixed $\Delta_{0}$ one finds three possible types of density driven changeovers
\cite{unpublished,czart01}: 
(i)  for $2\geq \Delta _0/D\geq 0$, "BCS"$\longrightarrow $  
Mixed $\longrightarrow $ "BCS";
(ii) for $\Delta_{0}/D>2$: "BCS" $\longrightarrow$ "LP" and 
(iii) for $\Delta_{0}/D<0$: "LP" $\longrightarrow$ "BCS".
Only if the LP level is
deeply located below the bottom of the $c$-band,
the system remains in the LP regime for any $n\leq 2$.

Let us  also comment on the effects of a weak interlayer coupling on the calculated
transition temperatures \cite{castro,hikami}.
In the KT theory the 2D correlation length behaves  as follows for
$T>T^{KT}$; $\xi(T)=a\exp\left(b/\sqrt{T/T^{KT}-1}\right)$, where 
$b \approx 1.5 $ and $a$ is the size of the vortex core. If 
 $U_{c}$ is the
coupling energy per unit length between the planes and $U_{c}\ll T^{KT}$, then  
the actual $T_{c}$ can be estimated by calculating the  energy needed
to destroy phase coherence between two regions of size $\sim \xi^2$ in different
planes i.e.  $T_{c}\sim {\bar c} U_{c}\left(\xi(T_{c})/a\right)^2$, where 
${\bar c}$ is the
interplanar distance. 
The resulting equations for $T_{c}$ can be solved asymptotically 
\begin{eqnarray}
T_{c}=T^{KT}\left(1+\frac{4b^2}{\ln^2(T^{KT}/{\bar c}U_{c})}\right),
\end{eqnarray}
therefore $T_{c}$  
%will grow with  interlayer coupling but 
is only weakly dependent on the interplanar distance ${\bar c}$ and is close to
$T^{KT}$, if $U_{c}\ll T_{KT}$.
In the presence of the interplanar coupling there is no discontinuous jump in
$\rho_{s}$ but a crossover from 2D like to 3D like (XY) behavior occurs.

\section{Superfluid transition from the pseudogap state}

 Let us now consider the pseudogap behavior and present the recent 
  evaluation of
  the superconducting 
  transition temperature from a pseudogap state by going beyond the BCS-MFA.
  In our analysis we have applied  a generalized $T$-matrix approach adapted to a
  two-component boson-fermion model\cite{rmunpub}. Our approach is an extension of the pairing
  fluctuation theory of the BCS-Bose-Einstein crossover \cite{zurich,levin}
  developed previously for a one-component fermion systems with attractive
  interaction.
  The numerical results  presented in Fig.3. are for a 3D sc  lattice assuming
  the tight-binding dispersion for fermions and bosons of the following form:
  $\epsilon_{\bf k} = D(1-{\tilde \gamma}_{\bf k}),~ D=zt$; ~~
$J_{\bf q}=J_{0}  {\tilde \gamma}_{\bf q},  ~J_{0}=zJ$,
 ${\tilde \gamma}_{\bf k}=\left[\cos(k_{x})+\cos(k_{y})+cos(k_{z})\right]/3$, $z=6$.

The results are shown for both cases with and without the direct 
hopping  of  LP's $J_{ij}$.
The calculated $T_{c}$'s are much lower as compared to 
BCS-MFA results (these are given by Eq.(\ref{tcmfa})), 
and if $J=0$, $T_{c} $ is strongly depressed  
as soon as the LP level is close to the bottom of the electronic band.
In the pseudogap region the electronic spectrum is gapped, and the pseudogap
parameter at $T_{c}$ for $\Delta_{0}>0$ essentially 
measures  a mean square amplitude of the
pairing field (of the "c" electrons).
The values of pseudogap parameter at $T_{c}$ 
are comparable to the zero temperature gap values  in the fermionic spectrum,
except for the c-regime. 

With the direct LP hopping $J_{0}/D=0.1$, which corresponds to $m_{B}=10 m_{F}$,
the hard-core bosons can undergo a  superfluid transition even without the
intersubsystem coupling
$|I_{0}|$. As we see from Fig.3 in the presence of the boson-fermion coupling 
$|I_{0}|$ the transition temperature is enhanced in the mixed regime.

In the self-consistent $T$-matrix approach  
%only 
the (amplitude) fluctuations of the order
parameter are included at the Gaussian level.
Nevertheless, it is interesting to observe 
that the phase diagram for $J_{0}=0$ shown in Fig.3 displays 
 similar  regimes as that of Fig.1 
determined in Sec.III from BCS and KT theories.

\section{Final Remarks} 
In conclusion, we summarize the important features of the model
considered\cite{rmsrbt,unpublished,physicac,jrmr96}.
\begin{enumerate}
\item Well defined Bogoliubov quasiparticles can exist in the superconducting 
ground state. However, above $T_{c}$ (in a mixed regime) local pairs coexist  
% together 
with itinerant fermions and the  normal state properties 
 deviate from Fermi liquid behavior. 
 
\item In the mixed regime, $T_{c}^{MFA} <T< T_{c}^{KT}$, 
the system will exhibit a {\underline {pseudogap}} in the $c$-electron spectrum, 
which will
evolve into a real gap as one moves to the LP regime.
For $\Delta_{0}<0$,   LP's  coexist with 
preformed $c$-electron pairs, which have the binding energy 
$E_{b}^{c}/2\propto T_{c}^{MFA}$.

\item For $d$-wave pairing, the superfluid density exhibits 
linear in T behavior at low T due to the presence of   
nodal quasiparticles. 
%The scaled superfluid stiffness
%$\rho_{s}(T)/\rho_{s}(T_{c})$ vs $T/T_{c}$ 
%shows only a weak dependence on the total density.

\item The Uemura-type plots i.e., the  $T_{c}$ 
{\it vs} zero-temperature phase stiffness 
$\rho_{s}(0)$,  are obtained for $d$, $s^{*}$ and $s$ -wave symmetry 
in the KT scenario \cite{unpublished}. The reason for Uemura scaling $T_{c}\sim \rho_{s}(0)$ is the 
 separation of the energy scales for the pairing
and for the phase coherence.\cite{rmsrbt,unpublished}.
\item The calculated  $T_{c}$'s in a 3D model beyond the BCS-MFA show crucial effects
of pair fluctuations in the mixed and LP regimes.
\end{enumerate}

 Some of our findings  can be qualitatively 
 related to experimental 
   results for the cuprate HTS where a pseudogap exists. It has
   been suggested  by ARPES experiments, that for underdoped cuprates 
   the Fermi surface in the 
   pseudogap phase is truncated around the corners 
   due to the formation of preformed (bosonic) pairs with charge $2e$, 
   whereas the "electrons" on the
   diagonals 
    remain  unpaired [7a].
   In the present two component model such a situation is obtained 
   when LP's and $c$-electrons coexist in the mixed regime 
   %(and $t_{2}/t<0$). 
   The linear $T$-dependence of the 
   superfluid density has been observed 
   experimentally in  copper oxides and also in 
   several organic superconductors. 
   This points 
   to an order parameter 
   of $d_{x^2-y^2}$-wave symmetry and existence of nodal quasiparticles. 
   In the present model the gap ratio is nonuniversal for all the pairing
   symmetries and can  deviate strongly 
   from BCS predictions (particularly in the d-wave case 
  for which it is  always  enhanced) \cite{unpublished}.  
   This feature  is also found in
   several  exotic superconductors.   
   The Uemura plots and the scaling $T_{c}\sim \rho_{s}(0)$ 
 reported for  cuprates and organic
   superconductors 
   can be reproduced within the model for 
   %both
   extended $s$- and $d$-wave order parameter symmetry 
   \cite{rmsrbt,unpublished}.
\section*{Acknowledgments}
 R.M. and S.R. acknowledge partial support 
from the State Committee for Scientific Research (KBN Poland): 
Project No: 2~P03B~154~22. R. M. also acknowledges support from the Foundation
for Polish Science.

%%%%%%%%%%%%%%%%%%%%%%%%%%%%%%%%%%%%%%%%%%%%%%%%%%%%%%%%%%%%%%%%%%%%%%%%
%%                            References
%%%%%%%%%%%%%%%%%%%%%%%%%%%%%%%%%%%%%%%%%%%%%%%%%%%%%%%%%%%%%%%%%%%%%%%%

\vfill

\newpage

\begin{figure}
\epsfxsize=12truecm
\centerline{\epsfbox{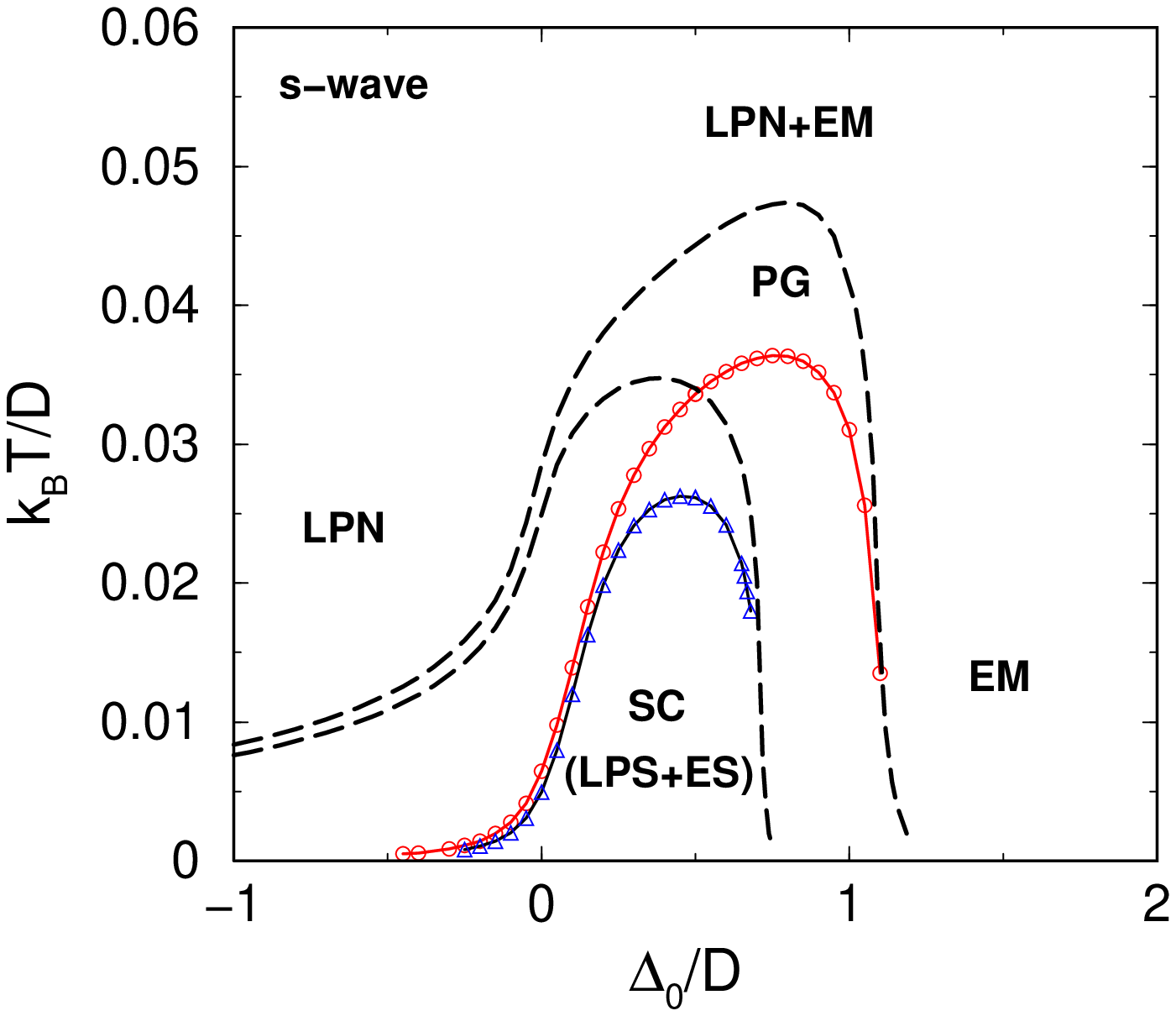}}
%\vspace{0.1cm}
\caption{{\small Phase diagrams of the induced pairing model as a function of
$\Delta_{0}/D$  at fixed $n$ derived for  $s$-wave symmetry. $J_{0}=0$  and $t_{2}=0$.
 $|I_{0}|/D=0.25$, D=4t. 
The dashed lines 
show the BCS-MFA transition temperature (upper for $n=1$ and lower for $n=0.5$), 
 while the lines with circles and triangles show the KT transition temperatures
calculated for $n=1$ and $n=0.5$, respectively.  
LPN--normal state of predominantly  LP's, EM--electronic metal, 
LPS+ES--superconducting (SC) state, PG -- pseudogap region. A weak interplanar
coupling stabilizes the SC state. }}
\end{figure}

\begin{figure}
\epsfxsize=12truecm
\centerline{\epsfbox{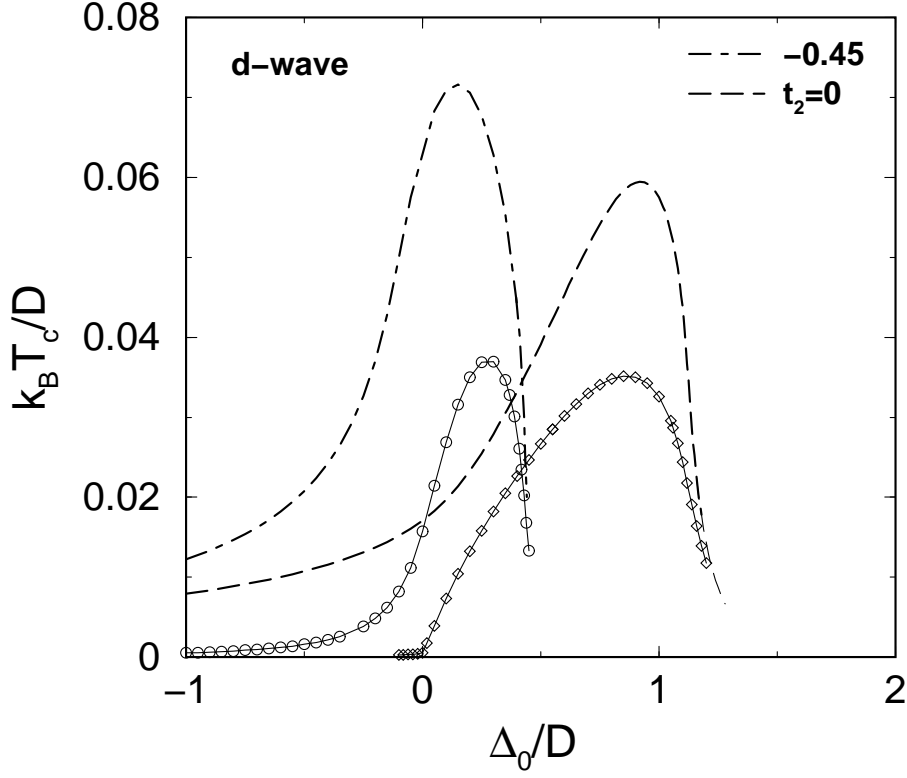}}
\caption{{\small MFA and KT transition temperatures   as a function of
$\Delta_{0}/D$ at fixed $n=1$ derived for  $d_{x^2-y^2}$ - pairing symmetry.  
The dashed and dot-dashed lines 
show the BCS-MFA transition
temperatures for $t_{2}=0$ and  for  $t_{2}/t=-0.45$, respectively.
 The line with diamonds shows the corresponding KT transition temperature
calculated for $t_{2}=0$ and the line with circles for $t_{2}/t=-0.45.$ 
$|I_{0}|/D=0.25$, $J_{0}=0$.}} 
\end{figure}

\begin{figure}
\epsfxsize=12truecm
\centerline{\epsfbox{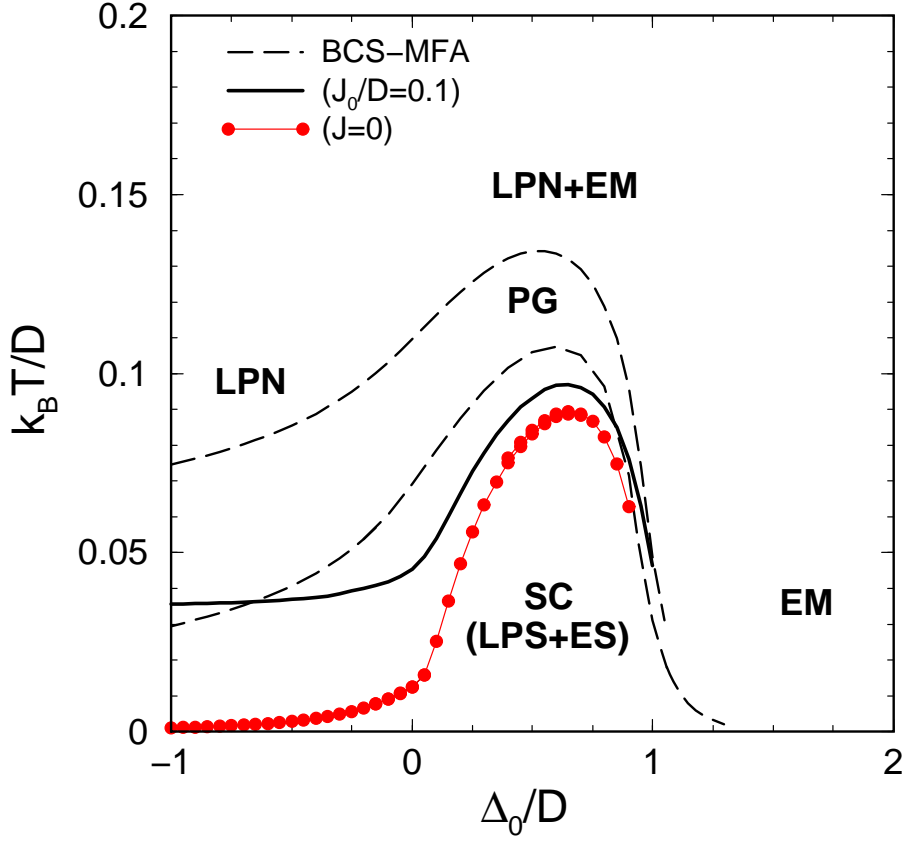}}
\caption{ {\small Phase diagrams of the hard-core boson-fermion model 
as a function of
$\Delta_{0}/D$ for  
$s$-wave pairing and sc lattice. $n=0.5, |I_{0}|/D=0.5, D=6t$.
 The transition 
temperatures derived with a $T$-matrix  approach
 are for  two values of $J$, which 
are shown by the
 solid line  ($J_{0}/D=0.1$) and the line with symbols ($J_{0}=0$), respectively.
The dashed lines indicate BCS-MFA transition 
temperatures (upper for $J_{0}/D=0.1$,
lower  for $J_{0}=0$).
LPN--normal state of predominantly  LP's, EM--electronic metal, 
LPS+ES--superconducting (SC) state, PG -- pseudogap region. }}
\end{figure}
\end{document}